\documentclass[jap,footinbib,soprop.cls ,11pt]{revtex4-1}
\usepackage{amsmath}    
\usepackage{graphicx}   
\usepackage{verbatim}   
\usepackage{color}      
\usepackage{subfigure}

\newcommand{\cM}{{\mathcal {M}}}
\newcommand{\mbf}[1]{\hbox{\boldmath $#1$}}
\newcommand{\mum}{\hbox{$\mu$m}}
\newcommand{\degree}{\hbox{$^\circ$}}

\newcommand{\eg}{\textit{e.g.}}
\newcommand{\ie}{\textit {i.e.}}
\newcommand{\etal}{\textit {et al.}}

\newcommand{\procspie}{\textit{Proc.\ SPIE}}

\begin{document}

\title{ {\rm Journal of Applied Physics 122, 055301 (published online 2 August 2017)}\break Quantum efficiency modeling for a thick back-illuminated astronomical CCD}

\author{D. E. Groom}\email{degroom@lbl.gov}
\author{S. Haque}
\author{S. E. Holland}
\author{W. F. Kolbe}

\affiliation{Lawrence Berkeley National Laboratory,
Berkeley, CA 94720, USA}

\date{\today}

\begin{abstract}
The quantum efficiency and reflectivity of thick, back-illuminated  CCD's being fabricated
at LBNL for astronomical applications are modeled and compared with experiment.
 The treatment differs from standard thin-film optics in that (a) absorption is permitted in any film, 
 (b) the 200--500~$\mu$m thick silicon  substrate is considered as a thin film in order to observe the fringing behavior at 
long wavelengths, and (c) by using approximate boundary conditions, absorption in the surface films 
is separated from absorption in the substrate.  
For the quantum efficiency measurements the CCD's are normally
operated as CCD's, usually at $\textrm T = -140\degree$C, and  at higher temperatures as photodiodes. They
are  mounted on mechanical substrates. Reflectivity
is measured on air-backed wafer samples at room temperature.  The agreement between  model expectation and
quantum efficiency measurement is in general satisfactory.  
\end{abstract}

\pacs{42.30Lr}

\maketitle



\section{Introduction}

\begin{figure}
\centerline{\includegraphics[scale=1.0]{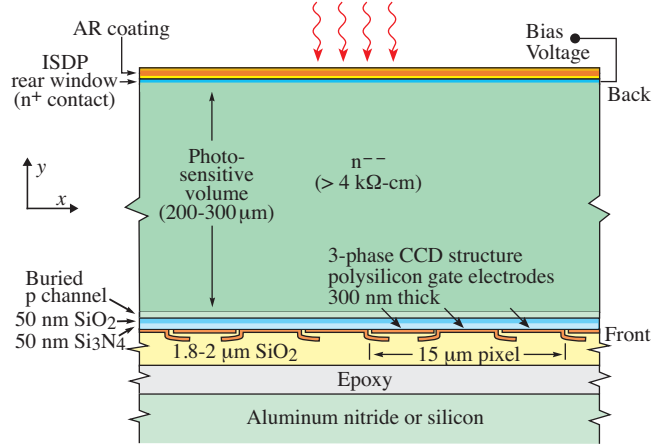}}
\caption{
Optical structure of the LBNL CCD.  In different versions the substrate resistivity ranges from 
4 to 20~k$\Omega$-cm, and some of the CCD's have a 10.5~\mum\ pixel width.   \textit{Not to scale}. 
\label{ccd-sim-inv}
}
\end{figure}

Fully depleted thick back-illuminated p-channel charge-coupled devices (CCD's) developed at 
Lawrence Berkeley National Laboratory  (LBNL) for astronomical applications have useful 
quantum efficiencies (QE's) extending into the near infrared (IR)\cite{holland03,HVccd06}.  
The response is limited to $\alt$1100~nm by the indirect bandgap of silicon.  
The QE  typically falls to about 50\% at  1000~nm, depending on the temperature and thickness  of the CCD.
Thicknesses range from 200 to 500\,\mum. 
The temperature range of interest is  $-140\degree$\,C to 20$\degree$~C.   
If $\textrm T\agt-100\degree$\,C, the QE is measured by operating the device as a photodiode. 
A highly schematic cross section of a typical back-illuminated LBNL CCD is shown in Fig.\,\ref{ccd-sim-inv}.  

Modeling the response depends crucially on  the complex refractive indices  and thicknesses 
of the materials involved. Especially critical is the absorption coefficient $\alpha(\lambda,T)$ 
of silicon as the indirect bandgap is approached.  Most of the other indices also present  special problems.

While the formalism presented  here is applicable to any CCD, we specialize to the LBNL case.
In these CCD's, a thin  film (10--25~nm) of {\textit{in-situ}} doped polysilicon
(ISDP) is grown on the rear surface to serve as an ohmic contact. Absorption in this layer limits 
the blue response, particularly below 450~nm.  Over the ISDP is an antireflective coating optimized for maximum
transmission at  desired wavelengths, particularly in the near-infrared.

This paper and the corresponding  code grew out  of  work reported by Groom\cite{spie99} in 1999.  
At that time the treatment of absorption in the ISDP and surface films was \textit{ad hoc} at best.
Absorption by an indium-tin oxide (ITO) film was neglected.  ISDP absorption was poorly modeled, 
so that results below about 550~nm were uncertain.

Although oblique incidence with either $\mbf{E}$ and $\mbf{B}$
parallel to the surface is treated, oblique incidence is of relatively little importance, since
even for the extreme case of an f1 system the incident $\cos\theta_0$ is $\ge0.89$, and the ray is 
``quickly straightened'' by refraction into the high-index silicon ($n=3.7$--4).

The  design constraints are daunting:  As shown in Fig.\,\ref{SiAbsLength}, the absorption length in 
silicon ranges over four orders of magnitude in the useful wavelength region,
from a few nm at the atmospheric cutoff near 320~nm to  the thickness of the CCD, typically 
250~\mum.  At the blue end of the spectrum it is close to the thickness of the (absorptive) ISDP, while at the near-IR end 
the CCD approaches transparency, with multiple reflections producing fringes.

\begin{figure}
\centerline{\includegraphics[scale=1.0]{Groom_Fig2%
.eps%
}}  
\caption{
The absorption length $\ell$ in silicon\footnote{The imaginary part of the index $k$, 
absorption coefficient $\alpha$, and absorption length $\ell$  
are related by $4\pi k/\lambda = \alpha = 1/\ell$.}.  The $300\degree$~K  solid curves are from the
\textit{Handbook of Optical Constants of Solids}\protect\cite{opticalhandbook85} extended to
1200~nm\protect\cite{janesickoptical}\ (black) and Green\protect\cite{green08} (gray or green).  
The dashed curves are 
calculated from the phenomenological fits by Rajkanan \etal\protect\cite{rajkanan79}.}
\label{SiAbsLength}
\end{figure}

The analysis uses the standard transfer matrix formalism described in multiple sources, \eg~in~Refs.\,\onlinecite{
pedrotti93,hecht,macleod}. Some departures are made in addressing CCD-specific absorption issues:
\begin{enumerate}

\item Many sources do not consider absorption---after all, one tries to make optical coatings out 
of transparent materials. But once it is introduced, there is a sign ambiguity in the definition of the complex index 
of refraction: $n_c = n+i\,k$ or $n_c=n-i\,k$.   The choice is arbitrary: The positive convention is used
in~Refs.\,\onlinecite{hecht,macleod,jenkinswhite50,panofsky-phillips,slaterfrank}, while the negative convention 
is used in~Refs.\,\onlinecite{born-wolf, pedrotti93,sommerfeld}. 
Having chosen the sign, one must take care that the rest of the formalism ensures that light is attenuated in
absorptive layers.  \textit{We adopt the negative sign convention.}

\item The silicon substrate  is also treated as a ``thin film,''  although it is opaque over much of the optical
range of interest. Fringes are commonly observed near the red end of a back-illuminated
CCD's sensitivity, implying coherence over the thickness of the device and, remarkably, 
near-specular reflection from the front surface gate structure and mechanical substrate.   
For a 200~$\mum$ thick CCD the fringe spacing at $\lambda = 1000$~nm
is only 0.7~nm, and in any case the fringes are ``washed out'' by a finite aperture.

\item It is useful to find absorption in the antireflective (AR) coating and, separately, in the ISDP. This is done by modifying 
boundary conditions, valid when the absorption length in silicon is small compared to the substrate thickness.

\end{enumerate}

We assume 100\% internal quantum
efficiency, \ie, every photon absorbed in the sensitive region produces a collected electron or hole. 

Model and measurement results are discussed for three antireflective coating designs: ITO/SiO$_2$,
ITO/ZrO$_2$/SiO$_2$, and TiO$_2$/SiO$_2$.  

\section{Multilayer reflected and transmitted amplitudes and intensities}\label{Eparallel}

In the simplest case, a plane wave in medium $a$ with real index $n_0$ (usually air or vacuum) is incident at an angle 
$\theta_0$ on a  film
with complex index $n_1$ and thickness $d_1$.  It exits into medium $b$ with index $n_s$.  With the definition 
$\gamma=(n/c)\cos\theta$ (appropriate if $\mbf{E}$ is parallel to the surface),  the boundary conditions relate the fields
at the two interfaces:
\begin{equation}
\left(\begin{array}{c}E_a\\  B_a\end{array}\right) =
 \left( \begin{array}{c c}
 \cos\delta_1 & \displaystyle\frac{i\,\sin\delta_1}{\gamma_1}\\
  i\,\gamma_1\sin\delta_1 & \cos\delta_1
  \end{array}   \right)
  \left(\begin{array}{c}E_b\\  B_b\end{array}\right)
\equiv \cM_1  \left(\begin{array}{c}E_b\\  B_b\end{array}\right)                                            
\label{matrix}
\end{equation}
The phase lag in one traversal, $\delta_1$, is $(2\pi d_1/\lambda)n_1 \cos \theta_1$. The product
$n_1 \cos \theta_1$ can be calculated from the complex version of Snell's law:
\begin{equation}
n_1 \sin \theta_1 = n_0 \sin \theta_0
\end{equation}

The transport matrix $\cM_1$ contains only variables pertaining to that layer; 
if the light is transmitted into another film a similar matrix is introduced.  For $N$ films,
\begin{equation}
\left(\begin{array}{c}E_a\\  B_a\end{array}\right)= \cM_1\cM_2\ldots\cM_N   
 \left(\begin{array}{c}E_b\\  B_b\end{array}\right)
\equiv \cM \left(\begin{array}{c}E_b\\  B_b\end{array}\right) \ .
\label{Mtotal}
\end{equation} 

The reflected fraction of the light $r$ and the  transmitted fraction $t$ can be extracted from the  boundary
condition equations,
\begin{subequations}
\begin{eqnarray}r&= \displaystyle\frac{\gamma_0 (m_{11} + \gamma_s m_{12})  - (m_{21}+\gamma_s m_{22}) }
                 {\gamma_0 (m_{11} + \gamma_s m_{12}) + (m_{21}+\gamma_s m_{22} )} \\
t&= \displaystyle\frac{2\gamma_0}
          {\gamma_0 (m_{11} + \gamma_s m_{12}) + (m_{21}+\gamma_s m_{22})} \ ,
\end{eqnarray}\label{r_n_t} 
\end{subequations}
where the $m_{ij}$ are the components of the product matrix $\cM$.

Over most of the spectral region of interest, the silicon substrate is essentially opaque.  More specifically, 
$\sin\delta$ and $\cos\delta$ both contain potentially large factors $\exp(2\pi d |k|/\lambda)$. 
While it is easy to block numerical overflows, we have found it convenient to factor out the divergent behavior:  
\begin{equation}
\cM_j = 
 e^{-\delta_{Ij}} \cM_j^F \label{factoredmatrix} 
\end{equation}
Here $\delta_{Ij}$ is the imaginary part of $\delta_I$, and, because of our negative sign convention for the imaginary 
part of indices, it is always negative.
 Eqs.~(\ref{r_n_t}) become
\begin{subequations}
\begin{eqnarray}
r&= &\displaystyle\frac{\gamma_0 (m_{11}^F + \gamma_s m_{12}^F) - (m_{21}^F+\gamma_s m_{22}^F) }
{\gamma_0( m_{11}^F + \gamma_s m_{12}^F) + (m_{21}^F+\gamma_s m_{22}^F)}\\
t&=& \displaystyle\frac{2  \gamma_0\exp{\left( \sum \delta_{Ij}\right) } }
          {(\gamma_0 m_{11}^F + \gamma_0\gamma_s m_{12}^F) 
          + (m_{21}^F+\gamma_s m_{22}^F)}\ .\label{amplitude}
\end{eqnarray}
\label{MFtotal}%
\end{subequations}
Since the exponential factors cancel in Eq.~(\ref{MFtotal}a), 
the reflected amplitude $r$ is calculable for any amount of absorption, while  the transmitted 
amplitude $t$ is (essentially) zero for high absorption.  

If $\mbf{B}$ is parallel to the surface, then $\gamma = (n/c)/\cos\theta$.  Since the phase lag $\delta$ is geometrical, it is 
not affected by polarization. However,  in Eqs.~(\ref{r_n_t}b) and ~(\ref{MFtotal}b),  $\gamma_0\ (=(n_0/c)\cos\theta_0)$ 
\textit{in the numerator} is replaced by $(n_0/c)/\cos\theta_s$.

The fractional reflected intensity $R$ is  $ |r|^2$.  
The fractional intensity of the light transmitted  into the mechanical substrate is
\begin{equation} 
T = \frac{\Re{(n_s\cos\theta_s)}}{n_0\cos\theta_0}|t|^2 \ .
 \label{finalT}
\end{equation}
This fraction is not \textit{per se} 
interesting, but if a  fraction $A$ of the light is absorbed in intermediate layers, then $A=1-R-T$. This is
the QE of the CCD plus $A_\mathrm{coat}$,  the fraction of the light absorbed in the surface films 
including the ISDP.


The desired result, the QE, is the absorbed fraction in the substrate alone. It is 
necessary to separate absorption in the substrate from  absorption in the complete coating and to separate
absorption in the IDSP from (possible) absorption in the AR layers. 

We can rewrite Eq.~(\ref{Mtotal}) as
\begin{equation}
\left(\begin{array}{c}E_a\\  B_a\end{array}\right)
 =\cM_\mathrm{AR}  \cM_\textrm{ISDP} \cM_\textrm{Si}
\left(\begin{array}{c}E_b\\  B_b\end{array}\right)
\equiv \cM_\mathrm {coat}\cM_\mathrm {Si}
\left(\begin{array}{c}E_b\\  B_b\end{array}\right)\ ,
\label{Mfactored}
\end{equation}
where $\cM_\mathrm{AR}$ is the product of the transfer matrices for  the AR coating films, $\cM_\mathrm{ISDP}$
is the transfer matrix for the ISDP coating, $\cM_\textrm{Si}$ is the matrix for the silicon substrate, and 
$\cM_\mathrm{coat} = \cM_\mathrm{AR}\cM_\textrm{ISDP} $.

One wishes to ``get inside the device''  and sample  the fields just after the light exits the AR coating, or after exiting 
the AR + ISDP coatings.   Unfortunately, light is reflected back into the AR coating at the substrate interface,
and we have not found an algebraic solution for the fields after the AR layers or after the AR layers plus the 
ISDP coating. However, an alternative method is quite accurate over almost the entire spectral
region.  It depends on two features of the problem:

\begin{enumerate}

\item Since no light is transmitted over most of the wavelength region ($\lambda \alt 900$~nm, depending on 
the CCD thickness), we can replace the device with just the surface layers on a semi-infinite silicon substrate, and
via Eqs.~(\ref{r_n_t})  or (\ref{MFtotal}) find the fraction of the light transmitted 
($T_{\mathrm {coat}}$) and reflected ($R_{\mathrm{coat}}$) by the ISDP + AR layers alone: 
\begin{equation}
\left(\begin{array}{c}E_a\\  B_a\end{array}\right)
=\cM_\textrm{AR}  \cM_\textrm{ISDP} 
\left(\begin{array}{c}E_b\\  B_b\end{array}\right)
\equiv \cM_\textrm{coat}
\left(\begin{array}{c}E_b\\  B_b\end{array}\right)
\label{Mcoat}
\end{equation}
Then the fraction of  light absorbed by the ISDP plus AR layers  ($A_\mathrm{coat}$) is 
$1-R_\mathrm{coat} - T_\mathrm{coat}$.  This can be subtracted from the absorption in the complete device
to find the QE:
\begin{equation}
\textrm{QE} = A-A_\textrm{coat}
\end{equation}
This is exact over most of the CCD, where the absorption length is small compared with the substrate thickness.

\item No serious error is introduced by lumping the Si and ISDP together as the semi-infinite silicon substrate,
then calculating the transmission and reflectivity of the AR  films alone.
The index difference between Si and the ISDP is very small: 
The reflectivity  is $<$0.04\% for $\lambda \agt 400$~nm and $< 0. 01$\% for $\lambda \agt 650$~nm.  
One thus obtains $A_ \mathrm{AR}$, the intensity fraction absorbed by the AR coating alone, essentially by 
the ITO.  Then $A_ \mathrm{ISDP}= A_\mathrm{coat}- A_\mathrm{AR}$.  
\end{enumerate}


\section{Refractive indices}\label{indices}%

\begin{figure}
\centerline{\includegraphics[scale=1.0]{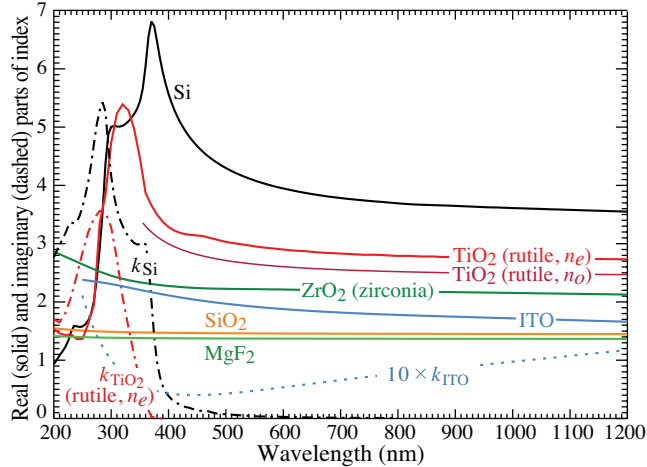}}
\caption{
Real and imaginary parts of the refractive indices of silicon and a few antireflective (AR) coating candidates.  
These are for bulk samples, and may not be realizable in sputtered or vapor deposited films.
The  ITO data are from SOPRA/ITO2.NK\cite{filmetrics} and are significantly different from those
used in our earlier work\cite{spie99}.}
\label{fig-indices}
\end{figure}

The indices of silicon and some candidate AR films are shown in Fig.\,\ref{fig-indices}.
Indices of the   transparent films (TiO$_2$, ZrO$_2$, HfO$_2$, 
fused SiO$_2$, MgF$_2$, etc.) are widely tabulated for bulk samples\cite{filmetrics}, but must 
be used with caution, especially for sputtered or vapor-deposited films.
For our application HfO$_2$ is inferior to ZrO$_2$. SiO$_2$ and MgF$_2$ have very similar 
properties, but since the LBNL MicroSystems Lab has wide experience with SiO$_2$ films,  
it  is used in preference to MgF$_2$.  With the exception of SiO$_2$, Si, ISDP, and all of the AR films used  present 
 special  problems that are discussed here.

More extensive studies of candidate materials have been published by Smith \& Baumeister\cite{smith79} 
and by Lesser\cite{lesser87}, focussing on UV transparent oxides and fluorides.

\subsection{Index of silicon}\label{siliconindex}%

While the coatings and boundary conditions at the surfaces of the CCD affect transmission and reflection,
it is absorption in the silicon substrate that results in charge collection. Understanding $k(T,\lambda)$ is
therefore  of paramount importance in understanding its QE.  This is particularly true in the near infrared, where the 
the absorption length $\ell\ (=1/\alpha=\lambda/4\pi k)$   rapidly approaches the thickness of the silicon 
substrate as the indirect bandgap  is approached
(at 1.12~eV, or 1100~nm, depending on  temperature), reducing the QE to nearly zero.

Many papers over the last 60 years have been devoted to the refractive index of silicon, particularly the absorptive
part, because of its great  importance in  solar cell design.  A subset of these
studies is particularly relevant to the optical and near-IR response of a
CCD\cite{green08,rajkanan79,macfarlane58,dash55,philip-taft60,
weakliem79,jellison82,jellison82B,jellison86,jellison92,buecher94,GreenKeevers95}.  Two of these are relevant
to our studies:

\begin{enumerate}

\item
Green\cite{green08} \hbox{(2008)} published tables of optical parameters at $300\degree$~K, 
together with empirical power-law temperature coefficients for $\alpha$, $n$, 
and $k$\footnote{To maintain the required significance at longer wavelengths it is necessary 
to recalculate $k$ from Green's absorption coefficients. Tables are most easily copied from the web version:
 \texttt{sciencedirect.com/science/article/pii/S0927024808002158\#} }%
: ``The self-consistent tabulation was derived from Kramers-Kronig analysis of updated reflectance data
deduced from the literature.''  Our earlier work\cite{spie99} used values of $n$ from the
\it Handbook of Optical Constants of
Solids\rm\cite{opticalhandbook85} extended to 1100~nm via tables presented by 
Janesick\cite{janesickoptical}. Green's values are quite close to these, but in our view supersede them.

Temperature coefficients  ``calculated from cited and additional data sets''  describe simple power laws as given 
in his Eqs.~(9) and (10).
Calculations using these coefficients indicate that the real part of the index varies only weakly with 
temperature. Changes are most evident near 380\,nm. Since Green's $n(300\degree$\,K) and $n(133\degree$\,K) 
are nearly indistinguishable, we have adopted Green's $n(300\degree$\,K) for  model calculations 
at all temperatures.

Absorption calculated using his coefficients yields model QE's seriously at variance with our data.  
An example is shown in
Fig.~\ref{DESI-data}.

\item
Rajkanan \etal\cite{rajkanan79} (1979)
developed a physics-based model of the absorption  that used 
 experimental data from MacFarlane \etal\cite{macfarlane58} and unpublished NASA sources 
to determine model
parameters\footnote{A nearly-identical expression for $\alpha$ is given by B{\"u}cher \etal\cite{buecher94} 
without reference to Rajkanan \etal's earlier paper. With two exceptions, the
parameters are the same or nearly the same.  However, the direct allowed bandgap contribution is  given as
$A(\hbar\omega-E_{gd})^{3/2}/\hbar\omega$, rather than Rajkanan \etal's $A(\hbar\omega-E_{gd})^{1/2}$.
Evidently the exponent should be 1/2 for a direct bandgap transition\cite{pankove71}.  Although both 
forms give similar results for photon energies below $E_{gd}$, we choose Rajkanan \etal's form.}.
The best accuracy was obtained ``with indirect band gaps at 1.1557 
and 2.5~eV and a direct allowed gap at 3.2~eV'' (390~nm).  

A fairly abrupt change in the 
absorption coefficient as the photon energy crosses this 
threshold (which increases somewhat with temperature) is evident in Fig.~\ref{SiAbsLength}.
Although their paper implies only 20\% accuracy,  we find remarkable agreement 
between the model predictions and our measured CCD QE at different temperatures and substrate 
thicknesses.     At wavelengths below 390~nm,  the Rajkanan \etal\ curves differ somewhat from measured 
values. This is due to the simple, smooth curve for $k$ obtained from the Rajkanan \etal\  model in the direct bandgap 
region. It makes little practical difference, since (a) our QE measurements extend down only to 320~nm, and 
(b) absorptions lengths are so short (a few 10's of nm) that the ISDP layer already confuses the issue.

Satisfactory agreement between QE modeled using the Rajhanan \etal's absorption 
coefficient\cite{rajkanan79} and data is obtained in all cases for the fiducial region in which the QE drops from 90\% to 20\% of its maximum, spanning a CCD thickness range from 200\,\mum\ to 500\,\mum\ and 
temperature range from 20\degree\,C to $-$140\degree\,C.   A surprising low-energy
``skirt'' at the higher temperatures extending the QE well above the indirect bandgap energy is 
discussed in connection with the TiO$_2$/SiO$_2$ coated CCD's.

\end{enumerate}


\subsection{Index of \textit{in-situ} doped polysilicon (ISDP)}\label{ISDPindex}%

During fabrication, a fairly thick layer of phosphorus-doped polycrystalline silicon on the rear surface of
the CCD acts as an active getter, maintaining the necessary very low leakage currents through
high-temperature processing steps.  
After thinning and polishing a backside ohmic contact is formed by depositing a thin ISDP layer (10--25~nm). 
The process is described in more detail in Ref.\,\onlinecite{holland03}.  Also shown in that paper is 
a secondary ion mass spectroscopy depth profile of a nominal 20~nm thick ISDP layer. The P concentration varies
by a factor of three until a depth of 20~nm is reached, then drops exponentially at about  one decade/7~nm.  It
is difficult to convert this profile to a single number for model calculations. In most cases the model thickness 
needs to be increased by 5--10 nm from the nominal value to obtain agreement with measurements.

The real part of its index is about the same as that of silicon, but it is considerably more absorptive than silicon
in the blue\cite{lubberts81}.  Tables SIPOLY$M$.NK in the SOPRA database\cite{filmetrics} provide 
$n$ and $k$  for $M$ = 10--90.  The peak values of both $n$ and $k$ decrease as $M$ increases.  
Documentation of the SOPRA tables is not available, but it is likely that $M$ is the fraction (in \%) 
of amorphous silicon present. For $\lambda>450$\ nm $n$ rises slightly with $M$, while $k$ increases significantly. 

\begin{figure}
\centerline{\includegraphics[scale=1.0]{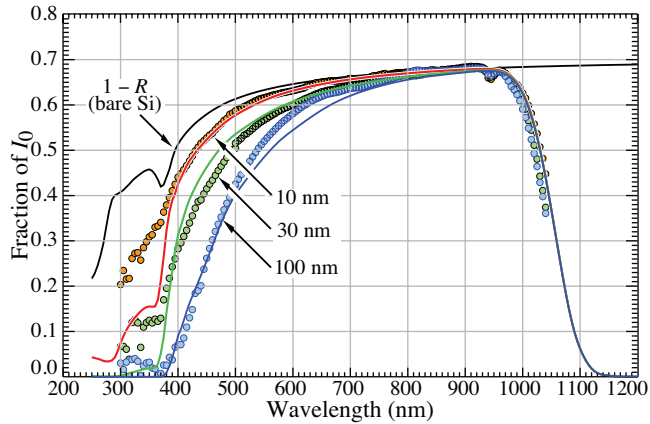}}
\caption{
Quantum efficiency of bare silicon with ISDP coatings 10, 30, and 100 nm thick.  Measurements
were made at ``room temperature,'' and the model calculations, using indices from the SOPRA SIPOLY10.NK table, 
were at 300\degree\,K.}
\label{ISDPmoses}
\end{figure}

Holland, Wang, and Moses fabricated and measured the QE of a series of photodiodes with successively thicker 
ISDP coatings\cite{HollandWangMoses97}.  A sample of their data together with model fits using  
SIPOLY10.NK  are shown in Fig.\,\ref{ISDPmoses} for ISDP 
thicknesses of 10, 30, and 100~nm.  In the 10~nm case the data and model are in essential agreement above
390~nm. The agreement is worse for the 30 and 100~nm cases, but the disagreement is in different 
directions.  In all cases, and in our CCD QE measurements,  the model QE falls below data for 
$\lambda\alt380\,$nm; 
our assumed index is simply too absorptive in this region. Since the QE is already falling rapidly here, the error
is of little practical consequence.

In the model QE calculations, ISDP absorption peaks at about 350 nm, and falls to insignificance in the red.
 Since there are yield concerns if the ISDP is too thin,
most recent CCD's use 25~nm coatings at the expense of significant QE loss below 500~nm.

\subsection{Index of indium-tin oxide (ITO)}\label{ITOindex}%

ITO films are (nearly) transparent to visible  and near-IR light.
In spite of ITO's widespread use, it is not an optically invariant material.  Its optical properties depend upon the method
of application, temperature, pressure, sputtering atmosphere and  power, composition, and annealing.  It varies
from amorphous to crystalline, and often has a graded 
index\cite{woollam94,gerfin96,lee93,rhaleb02,guillen07}.  

ITO was first used in our application to augment rear surface conductivity and also, with careful thickness choice, 
to act as an AR coating or first layer of an AR coating\cite{HollandWangMoses97}.  
It was reactively sputtered at room temperature in a low-pressure O$_2$/Ar atmosphere from a target composed of 
90\% In$_2$O$_3$ and 10\% SnO$_2$\ by weight.  To optimize conduction in the ITO, the oxygen 
content of the film (less than saturated) was controlled using deposition parameters described in 
Ref.~\onlinecite{HollandWangMoses97}.  Annealing in N$_2$ for an hour at 200\degree\,C substantially 
improved the transmittance.

\begin{figure}
\centerline{\includegraphics[scale=1.00]{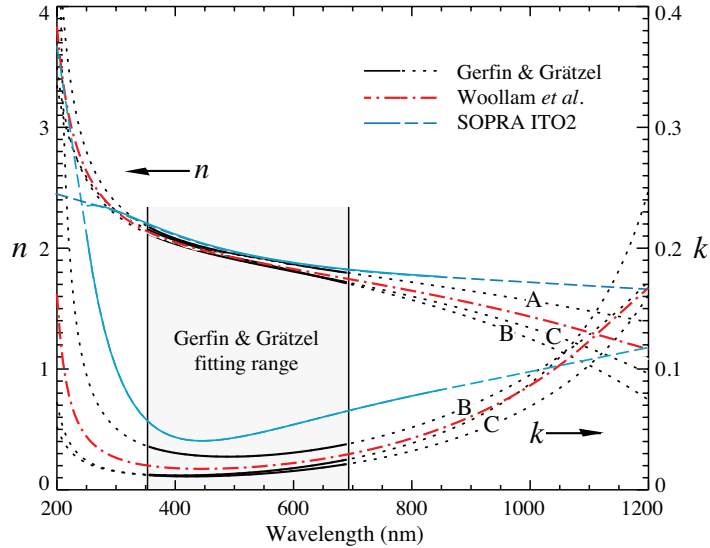}}
\caption{
Black curves show phenomenological fits to Gerfin \& Gr\"atzel's\cite{gerfin96}  spectroscopic 
ellipsometric measurements of ITO indices for samples A, B, and C between 1.8 and 3.5~eV
($\lambda\lambda = 689$--354~nm).  Dotted extrapolations were made using their dispersion formula. 
The dash-dotted red curves are functions of the same form drawn
through the measurements by Woollam \etal\cite{woollam94}.  Solid magenta curves are the ITO2 
indices from the SOPRA database (250--850~nm), with dashed linear extrapolations to 200--1200~nm.}
\label{ITO2sopra} 
\end{figure}

Our original sources of information about the optical constants for indium-tin oxide were the papers by 
Woollam, McGahan, \& Johs\cite{woollam94} and Gerfin \& Gr\"atzel\cite{gerfin96}.
Reference \onlinecite{gerfin96} gives tables of 6-parameter fits to a
dispersion formula for the dielectric constant~$\epsilon$. Figure~\ref{ITO2sopra}  shows the index, 
calculated as the real part of $\sqrt{\epsilon\mu_0}/c$, for three ITO films obtained from
different sources.  The Gerfin \& Gr\"atzel fits are based on data for $350<\lambda < 690$~nm.
Data from a figure in Ref.~\onlinecite{woollam94} as smoothed by Gerfin \& Gr\"atzel's dispersion 
formula  are shown by the dash-dotted red lines, the basis of our modeling calculations until recently.
The curvature change and decrease of the index above 700~nm are expected
from Drude-type absorption, indicating free carriers in the
film.  The solid magenta curves 250--850~nm are from ITO2.NK in the SOPRA database.
The SOPRA indices, linearly extrapolated to $\lambda = 1200$~nm,  have been used recently, and 
provide somewhat better agreement with the measured QE.  Since the dispersion fits show increasing slope 
of $k$ with $\lambda$ and decreasing slope of $n$ with $\lambda$, the SOPRA data 
and in particular our linear extrapolation may be unphysical. Attempts to fit a Gerfin \& Gr\"atzel-style dispersion function to 
the SOPRA data have failed to converge.  Index uncertainties in the extrapolated region
are relatively unimportant in our application.

\begin{figure}
\centerline{\includegraphics[scale=1.00]{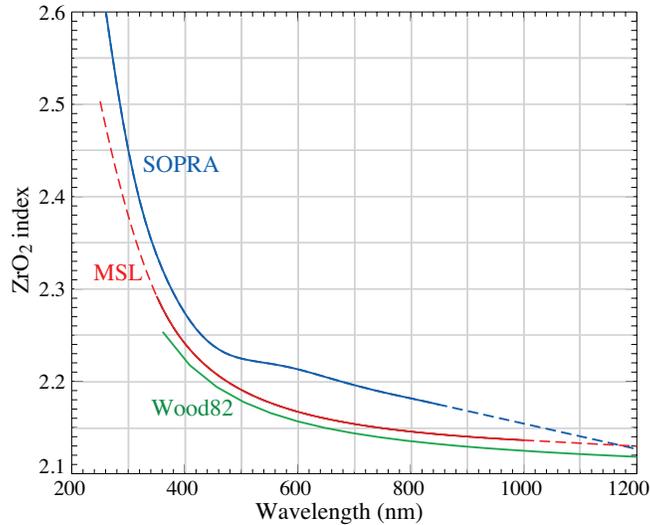}}
\caption{
The upper (blue) curve is the  ZrO$_2$ index of refraction as given in the SOPRA database 
(ZRO2.NK)\cite{filmetrics}.  The index shown in red (MSL, LBNL MicroSystems Lab) is 
from spectroscopic ellipsometer measurements made at the LBNL Molecular  Foundry.
The lower (green) curve is from Wood \& Nassau (1982)\cite{wood82}; their sample contained 12-mol \%
yttria.  Dashed curve segments indicate extrapolations.}
\label{ZrO2compare}  
\end{figure}

\subsection{Index of zirconium oxide}

The refractive index of ZrO$_2$ from three sources is shown in Fig.~\ref{ZrO2compare}.  The SOPRA 
data\cite{filmetrics} show a curious inflection at about 600~nm, perhaps from combining indices from 
different sources. The LBNL MicroSystems Lab (MSL) sputtered film spectroscopic ellipsometer measurements
were made at the LBNL Molecular Foundry.  The cubic zircona sample measured by Wood \& 
Nassau\cite{wood82} contained 12-mole~\% yttria. Since the index of Y$_2$O$_5$ is considerably lower than
that of ZrO$_2$\cite{nigara68}, one might expect the Wood \& Nassau sample to have a smaller index than
pure ZrO$_2$, as the figure suggests.

The MSL and Wood \& Nassau curves have roughly the same smooth shape, without the unexpected structure
near 600~nm.  We use the MSL index for our calculations.

\subsection{Index of titanium dioxide}

\begin{figure}
\centerline{\includegraphics[scale=1.00]{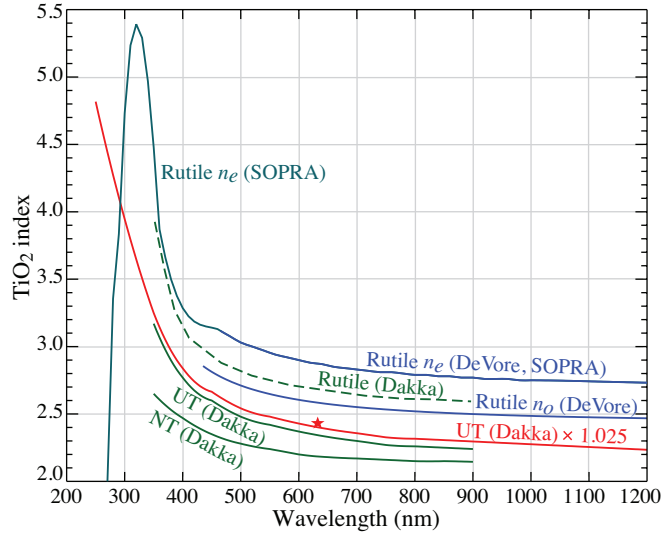}}
\caption{
TiO$_2$ refractive indices reported by Dakka \etal\cite{dakka99}, 
DeVore\cite{devore51}, and SOPRA\cite{filmetrics}.  ``NT'' indicates ``new target,''  
``UT'' indicates ``used target.''   The DeVore 1951 result for $n_e$ is evidently used in SOPRA 
TIO2.NK for $\lambda>436$\,nm.  The Dakka \etal\ UT result, extrapolated and scaled by 
1.025 (red curve), gives the best fit to our QE data.  The star indicates the measurement 
supplied by the manufacturer of the film, Hionix Inc.}
\label{TiO2-literature}

\end{figure}

The  index of the rutile form of TiO$_2$ shown in Fig.~\ref{fig-indices}  
(highest curve) is from SOPRA TIO2.NK, and tables are
available through Filmetrics\cite{filmetrics}. It may have the highest index for any transparent AR film candidate.
Its wavelength dependence is remarkably similar to that of silicon, making it the near-ideal material for a wideband
antireflective coating. But there are problems: the SOPRA TIO2.NK index is evidently for the extraordinary 
ray in this birefringent  material\cite{devore51,opticalhandbook85}.  
Moreover, reactively sputtered TiO$_2$ films at temperatures
consistent with CCD fabrication are mostly the (also birefringent) anatase form: ``Annealing of the films in air 
at 850~\degree C showed that anatase-rutile transformation strongly depends on the deposition temperature; 
the films deposited at temperature below 400\degree C were converted to the anatase-rutile mixture films, and the 
films deposited at 400\degree C to complete rutile films''\cite{wicaksana92}.  
The index of anatase is considerably below that of rutile, and tables are not readily available.  
The many papers on the subject do not present  consistent results\cite{dakka99,
tang94b,dannenberg00,jellison03,miao03}.  For example, Dakka \etal\cite{dakka99} show
different indices for ``new
target'' (NT)  and ``used target'' (UT) samples, and describe porosity and voids. The Dakka \etal\ results, together 
together with DeVore\cite{devore51} and SOPRA\cite{filmetrics} indices, are shown in 
Fig.\,\ref{TiO2-literature}.

\subsection{``Index" of the mechanical substrate}

When transmission becomes important in the near IR ($\agt900$~nm, depending on the
thickness of the silicon), the light exits into the gate structure and a mechanical substrate, which we have 
often modeled as exit into air: $n_s = 1.0$.   It is is  considerably more complicated than this, as shown in 
Fig.\,\ref{ccd-sim-inv}.  The light encounters 50~nm of SiO$_2$, then 50~nm 
of Si$_3$N$_4$, polysilicon gates 300~nm thick (with 40\% overlap), a thick layer of SiO$_2$, and epoxy that 
binds the device to thick silicon or aluminum nitride.  Remarkably, a single index $n_s$ seems to describe this
region adequately.  Modeled QE  turns out to be very insensitive to its value:
$n_s = 1.5$ or even $n_s=2.0$ moves the model calculation in the steep IR falloff region only slightly to the left,
where it agrees slightly better with the measurements.  However, asymptotic 1$-$$R$ and $T$ increase 
significantly with increasing~$n_s$. Examples are shown below for both $n_s=1.0$ and $n_s= 1.5$.


\section{Model calculations compared with QE and reflectivity measurements}\label{example}

\begin{figure}
\centerline{\includegraphics[scale=1.0]{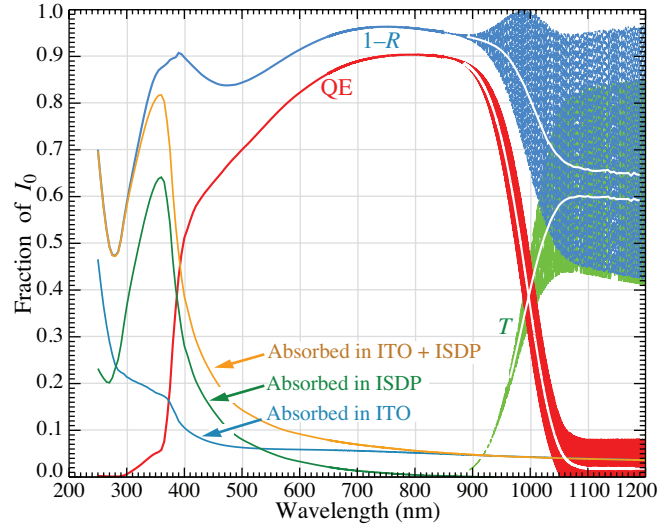}}
\caption{
Model calculation for a 200~\mum\ thick CCD 
at $-$140\degree~C with  80~nm SiO$_2$ and a 58~nm ITO AR coating. 
The ISDP layer is 25~nm thick.  The differences between the QE  and 1$-$$R$ at $\approx$\,800~nm and 
between the asymptotic $T$ and 1$-$$R$ are due to absorption in the  ITO.   Boxcar averages over the 
fringe bands  are indicated by the white curves. Here and in Fig.~\ref{CCDstd-data} the mechanical 
substrate index was taken as 1.0.  The approximations made in calculating 
coating absorption break down when there is appreciable transmission, producing the spurious ``foot'' of the QE 
for $\lambda\agt1100$~nm.}
\label{CCDstd-main}
\end{figure}

\begin{figure} 
 \centerline{\includegraphics[scale=1.00]{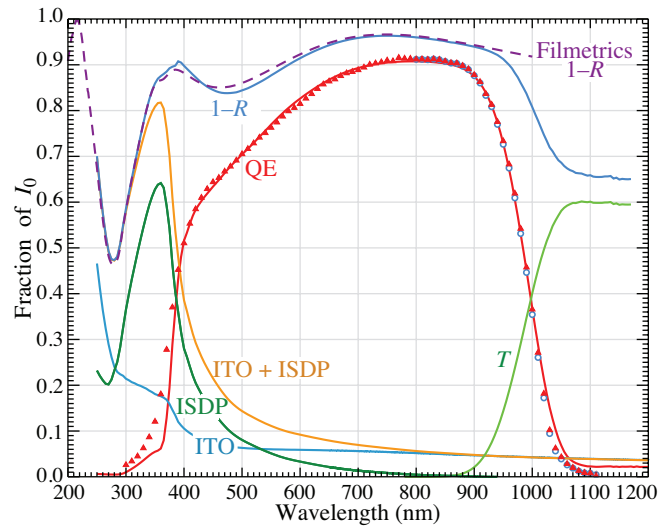}}
\caption{
Same as Fig.\,\ref{CCDstd-main} with fringe bands removed and  
QE measurements added.   
The blue circles on the descending part of the curve indicate the measured values  corrected for the 
monochromator bandwidth. The dashed 1$-$$R$ curve is from the online Filmetrics Reflectance
Calculator\cite{filmetrics}.
}
\label{CCDstd-data}
\end{figure}

Only the reflectivity $R$ and the transmitivity $T$ are directly modeled.  
As per the discussion of 
Sec.~\ref{Eparallel}, 
$1-R-T$ can be approximately decomposed into the QE, the light fraction absorbed in the ISDP layer 
($A_\mathrm{ISDP}$), and the fraction absorbed in the antireflective coatings  ($A_\mathrm{AR}$).  
If the AR coating is not significantly absorptive,  the QE and 1$-$$R$ curves should
be nearly identical for a fairly wide region in the red and near IR, where  there is no transmission and 
the ISDP is essentially transparent. 
If the AR coating does absorb significantly at all wavelengths, then there is a calculable gap between the 
QE and 1$-$$R$ curves, as is evident in Figs.~\ref{CCDstd-main} and \ref{CCDstd-data}.  
The measured QE is subject to amplifier gain uncertainties and other problems at the
few-percent level.  The absolute measurement of $R$ is used to normalize the QE measurements.   

The QE can be measured in either the normal CCD mode or by reading out the entire or a masked subsection
of the CCD as a photodiode (PD mode). This provides some additional redundancy, 
and measurements can be made at higher temperatures than are possible in the CCD mode.  An example
is discussed in Sec.\,\ref{hionixcoating}.

Our setup for measuring the QE\cite{QEmachine06} is fairly standard: light from a monochromator enters an
integrating sphere and exits a large aperture.  It arrives at the dewar containing the CCD after an 0.8~m drift space
in a baffled box. Slit widths are varied with light intensity; the bandwidth can be as small as 10~nm.
A room-temperature standard photodiode at the CCD's position and behind the same dewar window  
is used to calibrate a similar photodiode
at a small port in the integrating sphere that is then used as the reference for the QE measurement.

The reflectometer is described in Ref.~\onlinecite{reflectometer06}.  The intensity of a  light beam from the 
monochromator is measured by a photodiode after several mirror reflections. One mirror is then moved
so that a reflection from the (air-backed) CCD wafer sample is included in the optical path. The ratio yields the 
absolute reflectivity~$R$.  These measurements are at room temperature.  

The CCD's have from 4 to 16 readout amplifiers whose gain calibrations can be uncertain at the few percent
level.  In a broad red spectral band the QE should be nearly 1$-$$R$; this is used to normalize the QE measurements.

The CCD's fabricated and studied so far use various combinations of ISDP, ITO, ZrO$_2$, TiO$_2$, and SiO$_2$
films on the silicon substrate.  While the index of SiO$_2$ is very well known, the others all 
require special attention.  In particular, the near IR response of the CCD depends
crucially on the absorption coefficient of silicon, which is discussed in detail.

\begin{figure}
\centerline{\includegraphics[scale=1.0]{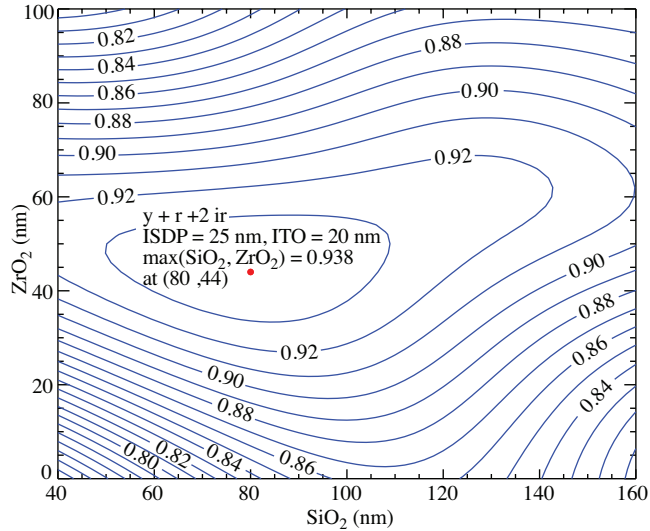}}
\caption{
Model  QE  contours as a function of ZrO$_2$ and SiO$_2$ thicknesses, 
where the QE is the average response of Hyper-Supreme-Cam \textit g and \textit i filters\cite{HSCfilters}.  The 
maximum shown, at ISDP:ITO:ZrO$_2$:SiO$_2$ = 10:20:106:42~nm, is close to the values chosen for the
DESI CCD's. The maximum of the model QE maximum using these values is 96.9\% at 875~nm.}
\label{DESIoptimized}
\end{figure}
\subsection{Two-layer AR coating design methodology}\label{design}

Recent designs use a minimal-thickness  ITO coating over the ISDP,  followed by high-index
and low-index layers. Although this is technically a 3-layer AR coating, the thickness of the ITO is held constant
in optimizing the other thicknesses for maximum response.

In the absence of any clear criterion for AR coating optimization, the response was maximized 
for several linear combinations of broadband and narrowband filter responses. 
Figure~\ref{DESIoptimized} shows an example for an 
ITO (fixed 20~nm thick)/ZrO$_2$/SiO$_2$ coating, where the model QE is the average of HSC-\textit g and HSC-\textit i
responses\cite{HSCfilters}.  Response was calculated for a matrix of 
ZrO$_2$ and SiO$_2$ thicknesses to make 
the QE contour plot.  Thickness tolerances can be estimated from the ``flatness'' of the peak.

Although still somewhat subjective, the method has been moderately successful.
Subsequent thickness tuning on the basis of
experiment has also been useful.  One problem, still not understood, is a discrepancy of 10--15\% between the
deposited SiO$_2$ thickness and the modeled thickness---the CCD behaves as though excessive SiO$_2$ 
has been deposited\footnote{Blacksberg \etal\ report measured film indices somewhat different than were expected\cite{blacksberg08}.
For SiO$_2$ they find a film index about 2\% higher
than for bulk fused quartz, in the right direction but not enough to explain our apparent excess.}
.  We correct for this empirically by depositing a thinner SiO$_2$ layer.

\subsection{ITO/SiO$_2$ coating}\label{stdcoating}

In initial designs, indium-tin oxide was used both to ensure adequate rear-surface conductivity and to serve as 
the first layer of the AR coating.  The model output is shown in Figs.~\ref{CCDstd-main}, 
\ref{CCDstd-data}, and~\ref{CCDstd-fringes}.  The full calculation is shown in Fig.~\ref{CCDstd-main}.  
In this and in most other examples, light is normally incident on the CCD.  
The decomposition of the absorption into QE, absorption in the ISDP+AR coating, ISDP, and AR
is discussed in Sec.~\ref{Eparallel}.   Above  900~nm the CCD becomes increasingly 
transparent, resulting in interference between the reflected and transmitted amplitudes in the silicon substrate.
The responses change rapidly with increasing wavelength, and appear as bands in Fig.~\ref{CCDstd-main}.
White lines show box-car averaged $T$, 
1$-$$R$, and  QE, the  intensity fraction absorbed in the sensitive region.  The discrepancy between 
1$-$$R$ and the QE in the near-IR is primarily caused
by absorption in the ITO.  The approximations made in separating the QE and absorption in the ISDP and ITO 
break down as fringing becomes more pronounced, so the modeled QE does not quite go to zero at long 
wavelengths.

Figure~\ref{CCDstd-data} shows the same curves, but with the fringing bands replaced by boxcar
averages, experimental QE 
measurements  (red triangles) added, and a dashed curve comparing the modeled 1$-$$R$ with 
results from the Filmetrics Reflectivity Calculator\cite{filmetrics} has been added.  
Blue circles on the descending part of the QE response illustrate a correction for the finite monochromator
bandpass.  

For this model calculation $n_s$, the effective index of the mechanical substrate, $n_s$, was taken as 1.00.
The value is relevant only in the transparency region, $\lambda \agt900$~nm.  

CCD's with this coating  have been used in the BOSS\cite{BOSSspectrograph} camera,  in DECam\cite{DECam},
in the red leg of the KECK Low Resolution Spectrometer\cite{LRISredleg}, and in other applications.

\subsection{(ITO)/ZrO$_2$/SiO$_2$ coating}\label{DESIcoating}

\begin{figure}
\centerline{\includegraphics[scale=1.00]{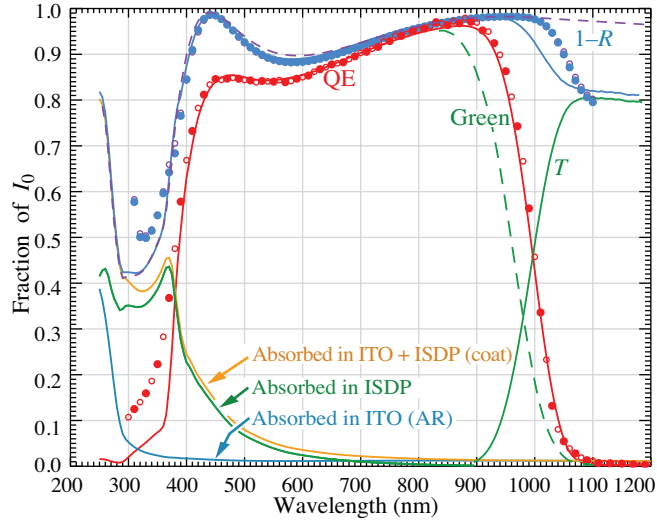}}
\caption{
The nominal ISDP:ITO:ZrO$_2$:SiO$_2$ thickness were 11:20:38:106~nm, while a better fit (shown)
is for thicknesses 18:20:38:118 nm. The wafer was 
250~\mum\ thick,   the temperature was  $-$140\degree\,C (model and QE data) and the mechanical substrate
index $n_s$ was 1.5.  Reflectivity was measured at room temperature.
Fringing bands  are deleted from the figure. Curves show the model results; points show  
measurements. QE data for 12 channels are scaled 
for agreement near the model QE peak.    The dashed curve is calculated using Green's
temperature coefficients\cite{green08}. }
\label{DESI-data}   
\end{figure}

A design has been developed for the Dark Energy Spectroscopic Instrument (DESI)\cite{DESICCD} that uses
ZrO$_2$ rather than ITO for most of the high-index layer. There is  a  20--25~nm ITO film  between the 
ISDP and ZrO$_2$.  The ITO is included to avoid direct contact with a new material that might
introduce reliability issues. (We have had substantial experience with ITO in direct contact with ISDP.)  
In any case, the ITO does not seriously compromise the QE in the DESI red and 
IR channels\cite{DESICCD}.  An example is shown in Fig.~\ref{DESI-data}. Compared with the ITO/SiO$_2$ 
response (Figs.~\ref{CCDstd-main} and \ref{CCDstd-data}), the blue QE is increased 
and flattened, and the QE is only slightly below the 1$-$$R$ limit in a broad red region.

For the model calculation shown here, $n_s$, the effective index of the mechanical substrate, was taken as 1.50.
A comparison with Fig.~\ref{CCDstd-main}, where it was 1.00, shows that the asymptotic 
transparency and 1$-$$R$ are both greater, although their  difference (nearly the QE in this case) is almost unchanged.

The near-IR QE using Green's $k(\lambda,-140\degree\,\rm C)$\cite{green08}  (Sec.~\ref{siliconindex})
is also shown in Fig.~\ref{DESI-data}.  
Green's agreement with Rajkanan \etal\ is better when the comparison is made at $T=300\degree\,$C.

\subsection{TiO$_2$/SiO$_2$ coating}\label{hionixcoating}

Since the LBNL MicroSystems Lab did not have a sputtering target for TiO$_2$ deposition, 
a TiO$_2$/SiO$_2$ coating was applied 
to an otherwise-complete CCD by the Hionix corporation\cite{hionix}.  
They reported a TiO$_2$ index of 2.44 at 633~nm.  The SiO$_2$ film was about 14\% thicker 
than requested, resulting in a QE somewhat lower than expected.

\begin{figure}
\centerline{\includegraphics[scale=1.00]{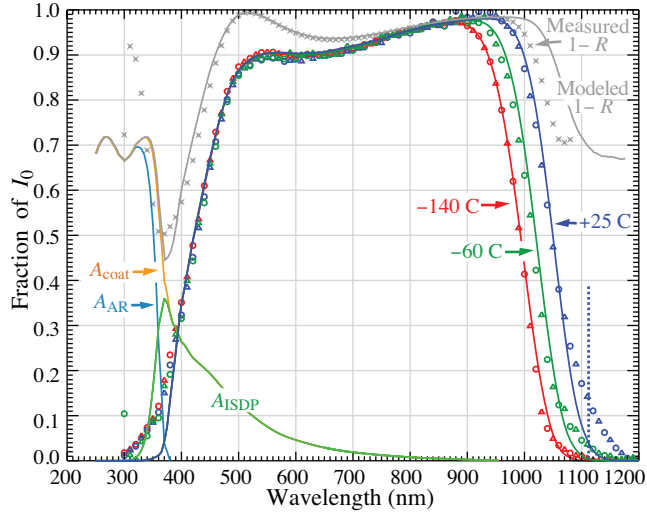}}
\caption{
Response of an ISDP/TiO$_2$/SiO$_2$ (32\,nm/57\,nm/132\,nm) coated CCD at $-140$\,C (CCD mode), 
$-60$\,C (PD mode) and 25\,C (PD mode).  Curves are the model results; points are the 
measurements.  Substrate index is 1.5 for model QE and 1.0 for model 1$-$$R$.  
A dotted vertical line at 1116\,nm indicates the indirect bandgap at 300$\degree$\,K.  
At the higher temperatures, particularly at 25\,C,  the QE does not fall to zero as rapidly as the model predicts,
evidently due to e-h production via a photon-two phonon processes\cite{pankove71}.
}
\label{TempCompare}
\end{figure}

The response is shown in Fig.\,\ref{TempCompare}.  Good fits to the measured QE were obtained using
the Dakka ``UT'' index scaled by 1.025, in agreement with the Hionix measurement at 633~nm.
This CCD has a better response than any of the others we have tested. Nonetheless, given the difficulty and 
likely  unpredictability of TiO$_2$ films and only marginal improvement from the 
(ITO)/ZrO$_2$/SiO$_2$ AR coating, there is little incentive to pursue this approach.

A remarkable feature of the room-temperature measurements is the QE ``skirt'' 
extending well beyond 1116~nm, the wavelength corresponding to the indirect bandgap at 1.1108~eV
at 300\degree K (vertical dotted line in Fig.\,\ref{TempCompare}). In order to conserve both energy and momentum, absorption involving indirect transitions
requires the absorption or emission of one or more phonons. The 
Rajkanan \etal\  model\cite{rajkanan79} takes into account the two lowest phonon excitations, 0.0183~eV
and 0.0577~eV that produce contributions to the absorption coefficient with displacements
of $\pm 19$\,nm and $\pm61$\,nm.  As a result, the
model QE  at 25\degree~C shown in Fig.\,\ref{TempCompare} is about 8\% at 1100 and falls to 
zero by 1177~nm.  However, the measured response is above 10\% at 1120~nm 
and  falls to zero only just below 1200~nm.  This  e-h production by low-energy photons is  thought to be due to 
higher-temperature double-phonon processes producing offsets of $\pm 99$~nm 
and $\pm 149$~nm\cite{macfarlane58}.  

\begin{figure}
\centerline{\includegraphics[scale=1.00]{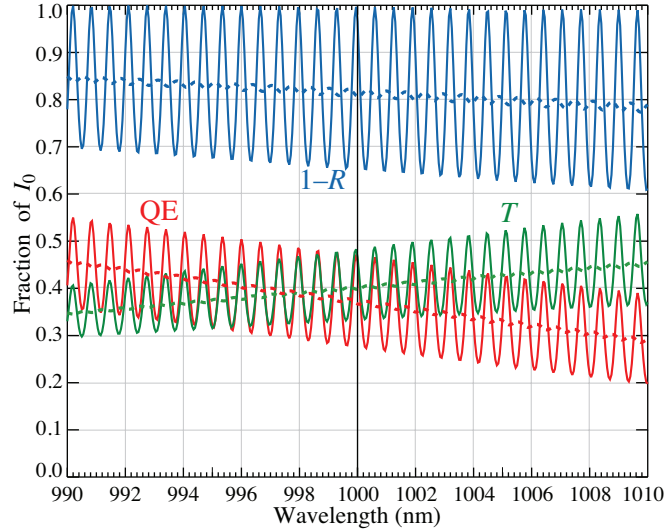}}
\caption{
Detail of the 990--1010~nm region of Fig.\,\ref{CCDstd-main}.  
The fringe spacing is $\delta\lambda = \lambda^2/2nd$, where $d$ is the silicon thickness and $n$ is the
real part of its index (at 1000~nm, $n = 3.57$ and $\delta\lambda=0.7$~nm). The dashed central curves
show the results of averaging the light's incidence angles over an f1.5 aperture.  
These are close to the boxcar averages,
which appear to be ``below center'' because of the asymmetry of the Haidinger fringes. }
\label{CCDstd-fringes}
\end{figure}

\subsection{Fringing}\label{fringing}

The sensitive region of a ``traditional'' thin back-illuminated CCD is an epitaxial layer $\approx\,$20~\mum\ thick. At longer
wavelengths light reaches the back surface, and the resulting multiple reflections  produce interference ``fringes.''  
A white-light exposure with the KECK low-resolution spectrograph 
(LRIS)\cite{oke95} is shown in Ref.~\onlinecite{spie99}.
The fringes are somewhat irregular because of thickness variations in the epitaxial layer, but one can 
infer from the fringe average spacing $\delta\lambda$ that the layer is 20--22~\mum\ thick. (For a thickness $d$ at
a wavelength where the real part of the silicon index is $n$, $\delta\lambda=\lambda^2/2nd$.) In this case, 
the spacing is about 7~nm at $\lambda=1000$~nm, providing at once a way to measure the epitaxial layer 
thickness\cite{janesick01} and a nuisance for observers. Broadband photometry using 
R and IR filters is plagued by swirled interference patterns whose amplitude is modulated by time-varying 
OH sky lines.  Removal of these fringes is discussed by McLean\cite{mclean97} and  others, and 
algorithms to treat the problem are part of observatory software.  That the fringes exist means that the light remains 
coherent over at least 20~\mum\ and that reflection from the mechanical substrate is remarkably specular.

An expansion of our modeled  990--1100~nm region for a 200~\mum\ thick CCD's is shown in 
Fig.~\ref{CCDstd-fringes}.    For normally incident light 
the peak-to-valley QE variation is nearly 0.2 of the incident intensity, or 0.34 of the average QE.  However,
the fringes are ten times closer than those in a 20~\mum\ thick CCD, or about 0.7~nm at 1000~nm. Only the
highest-resolution spectrometers could resolve these fringes.  In particular, they are not resolved in
our QE measurements because of the much-wider  bandpass
of the monochromator.  Even if they are not observed, the agreement of
the boxcar averaged QE with measured QE  argues for coherence and fairly specular reflection.

An actual instrument has a finite aperture, so that light arrives at different angles, each with a 
fringe spacing characteristic of the slant depth.  The fringing pattern tends to average out. The central dashed
lines in Fig.~\ref{CCDstd-fringes} were calculated for an f1.5 aperture. The remaining oscillations 
are the result of a beat pattern characteristic of this wavelength range; in the model
they disappeared and reappeared as the aperture was changed.  Results of the calculation are nearly 
independent of polarization.

Using a DECam CCD, Stevenson \etal\cite{stevenson16} searched 
for water in the atmosphere of an exoplanet with the LDSS-3 spectrometer (focal ratio f2.5) 
at the Magellan telescope at Las Campanas Observatory.  Fringes were not observed. A more detailed study of
one of the images by A. Seifahrt\cite{seifahrt16} set limits on the fringe intensity of 0.25\% (peak-to-valley).

DECcam Y-band images have shown fringing swirls at the 0.4\% peak-to-valley level\cite{martini12}.  
The pattern and spacing is probably inconsistent with interference in the CCD itself, given its thickness and  expected 
thickness variations.  It is possible that the interference originates in the 
epoxy layer binding the CCD to the mechanical  substrate, which was  aluminum nitride for the DECam CCD's. 

In any case, the steep QE falloff in the near IR limits any possible fringing  to  the 900--1050~nm region.
The fringing amplitude, especially in the cases of the DESI 
(ITO)/ZrO$_2$/SiO$_2$ CCD(Fig.~\ref{DESI-data})  and the TiO$_2$/SiO$_2$ 
CCD (Fig.~\ref{TempCompare}), is minimized by the very low reflectivity in this region.  
For normally incident light the maximum peak-to-valley QE variation is 0.10 for the DESI
CCD and 0.08 for the TiO$_2$/SiO$_2$ CCD.

\section{Prospects for QE improvement}

It was mentioned above that the model thicknesses had to be ``tuned'' somewhat to obtain agreement 
with the measurements.
To some extent this was because of variation in the actual film thicknesses, or, more likely, that the thin film indices were
different from those measured in bulk samples.  (In the case of TiO$_2$, there was even ambiguity about the 
crystal form.)   Thus by perturbing the actual film thicknesses the QE can be improved. 
Experimental AR coatings are being made to explore this.  It is likely that response of the (ITO)/ZrO$_2$/SiO$_2$ 
(DESI) CCD can be enhanced in the 450--600~nm region from about 84\% to about 90\%  by increasing 
the design ZrO$_2$ thickness by  7--10~nm---a spectral region just below the DESI red channel,
but the increased blue response might be useful for other applications.

Our blue response is limited by absorption in the ISDP rear electrode.  An alternate approach is delta doping, depositing
approximately a monolayer of dopant atoms on the rear surface via molecular beam epitaxy (MBE).   This technique 
has been developed at the Jet Propulsion Lab (Caltech) for a variety of spacecraft and sounding rocket 
missions\cite{nikzad16}.  The object is to obtain reflection-limited ultraviolet quantum efficiency down to about 100~nm.

Antimony layers about 5~nm thick were applied in this way to 2k$\times$4k and 1k$\times$1k  LBNL CCD's 
similar to those described here\cite{blacksberg08,jaquot10}.  QE measurements with and without a (nominally)
Si$_3$N$_4$/SiO$_2$ AR coating are 
shown in Ref.\,\onlinecite{blacksberg08}.  The results are consistent with ours except that their QE remains high down to
about 400~nm.

There has been no further work on  the delta-doping approach.

\section{Conclusions}

We have extended standard thin-film optics methods to model the quantum efficiency (QE) of LBNL thick, back-illuminated 
CCD's, and compared the results with experimental measurements of QE and reflectivity for three antireflective coating examples. The calculations included (a) considering the thick substrate as one of the films and (b) separating absorption
in  an ISDP coating serving as the rear contact and in the antireflective  coating from absorption in the 
silicon substrate (the QE) by modifying boundary conditions.  Problems 
encountered with  the indices involved are discussed.  While
agreement with experimental QE and reflectivity measurements are regarded as adequate, it is limited by 
uncertainties in the ISDP index, the ITO and TiO$_2$ indices, and film deposition thickness variations.
Fringing is neither expected nor observed.

    
\section{Acknowledgements}
The QE ``machine'' was built by Jens Stecker and the reflectometer by Maximilian Fabricius, both in collaboration
with and under the supervision of Armin Karchar.  With the exception of TiO$_2$, AR coatings were applied 
in the LBNL MicroSystems Lab.  We are also grateful for Chris Bebek's  encouragement and guidance.
{\color{red}.}

This work was supported by the U.S. Department of Energy under Contract No.\ DE-AC02-05CH11231.

\bibliography{qe_17_revtex}

\begin{thebibliography}{65}
\bibitem [{\citenamefont {Janesick}(iven)}]{janesickoptical}%
  \BibitemOpen
  \bibfield  {author} {\bibinfo {author} {\bibfnamefont {J.~R.}\ \bibnamefont
  {Janesick}},\ }\href@noop {} {\bibfield  {journal} {\bibinfo  {journal}
  {IS\&T/SPIE Symposium on Electronic Imaging Science and Technology}\ ,\
  \bibinfo {pages} {233}} (\bibinfo {year} {no data source or year
  given})}\BibitemShut {NoStop}%
\end{thebibliography}
\end{document}